\documentclass[letterpaper
]{quantumarticle}
\pdfoutput=1
\usepackage{subcaption}
\usepackage{capt-of} 
\usepackage{graphicx}
\usepackage{color}
\usepackage{hyperref}
\usepackage{float}
\usepackage{array, multirow}
\usepackage{amsmath, dsfont}
\usepackage{amsfonts}
\usepackage{newtxtext}
\usepackage{booktabs}
\usepackage{xcolor}
\usepackage{soul}
\usepackage{url}
\usepackage{physics}
\usepackage{graphicx}
\usepackage{color}
\usepackage{hyperref}
\usepackage{float}
\usepackage{array, multirow}
\usepackage{amsmath}
\usepackage{newtxtext}              \usepackage{booktabs}
\usepackage{xcolor} 
\usepackage{soul}
\usepackage{url}
\usepackage{qcircuit}
\usepackage[margin=0.8in]{geometry}

\hypersetup{colorlinks, 
    linkcolor={blue!75!black!80!yellow},
    citecolor={blue!75!black!80!yellow}, 
    urlcolor={blue!75!black!80!yellow}
    }

\begin{document}

\title{Quantum Simulation of Two-Level $PT$-Symmetric Systems Using Hermitian Hamiltonians}

\author{Maryam Abbasi}
\affiliation{Department of Chemistry, University of Minnesota, Minneapolis, MN 55455 USA}

\author{Koray Aydo\u{g}an}
\affiliation{School of Physics and Astronomy, University of Minnesota, Minneapolis, MN 55455 USA}
\author{Anthony W. Schlimgen}
\affiliation{Department of Chemistry, University of Minnesota, Minneapolis, MN 55455 USA}

\author{Kade Head-Marsden}
\affiliation{Department of Chemistry, University of Minnesota, Minneapolis, MN 55455 USA}
\affiliation{School of Physics and Astronomy, University of Minnesota, Minneapolis, MN 55455 USA}
\email{khm@umn.edu}


\begin{abstract}
Parity-time ($PT$)-symmetric Hamiltonians exhibit non-unitary dynamical evolution while maintaining real spectra, and offer unique approaches to quantum sensing and entanglement generation. Here we present a method for simulating the quantum dynamics of $PT$-symmetric systems on unitary-gate-based quantum computers by leveraging Hermitian equivalents through similarity transformations. We introduce two algorithms for simulating $PT$-symmetric systems near exceptional points where eigenvalues and eigenstates coalesce. The first is a hybrid classical-quantum algorithm and the second reduces the classical component by using an ancilla qubit. We then use perturbation theory to extend this method to consider the dynamics of two weakly interacting $PT$-symmetric systems. Demonstrations on a quantum device and noisy simulators validate the algorithm on current quantum devices, offering potential for scalable simulations of $PT$-symmetric systems and pseudo-Hermitian operators in quantum computation.
\end{abstract}

\maketitle

\section{Introduction}
\label{section:intro}
 Understanding the evolution of open quantum systems (OQSs), quantum systems which interact with an environment, is crucial for fundamental research and technological advancements in quantum information and sensing~\cite{Breuer2007, Kraus2008, Diehl2008, Verstraete2009, HeadMarsdenFlick2021, Olivera-Atencio2023}. Simulating these systems efficiently on quantum computers remains a challenge due to their non-unitary time propagators, which cannot be mapped directly into the unitary framework common in current quantum computers. Recent progress in open quantum simulation offers new approaches to address this challenge with different methods including imaginary time evolution~\cite{McArdle2019, Kamakari2022, Shirakawa2021, Matsushita2021, Pollmann2021} and block encodings~\cite{Hu2020, Schlimgen2021, Schlimgen2022a, Gaikwad2022, Suri2023, Ding2024, Basile2024, Xuereb2023}, among others~\cite{Gupta2020b, Berreiro2011, Wei2016, Mostame2017, Garcia-Perez2020, Patsch2020, Su2020, Sun2024, Burger2022, Wang2022, Cattaneo2023, Zanetti2023, Leppaekangas2023, Guimaraes2023, Guimares2024, DelRe:2024, Childs2017, Li2023, Li2023a, Cosacchi2018, Liu2019, Cygorek2022, Andreadakis2023, Regent2023,Muller2011, Ramusat2021, Mahdian2020b, Wang2023c, Dutta:2024, DelgadoGranados2025}.

\begin{figure}[ht]
    \centering
    \includegraphics[width = 1\columnwidth, trim = 0 15.5cm 7cm 0, clip]{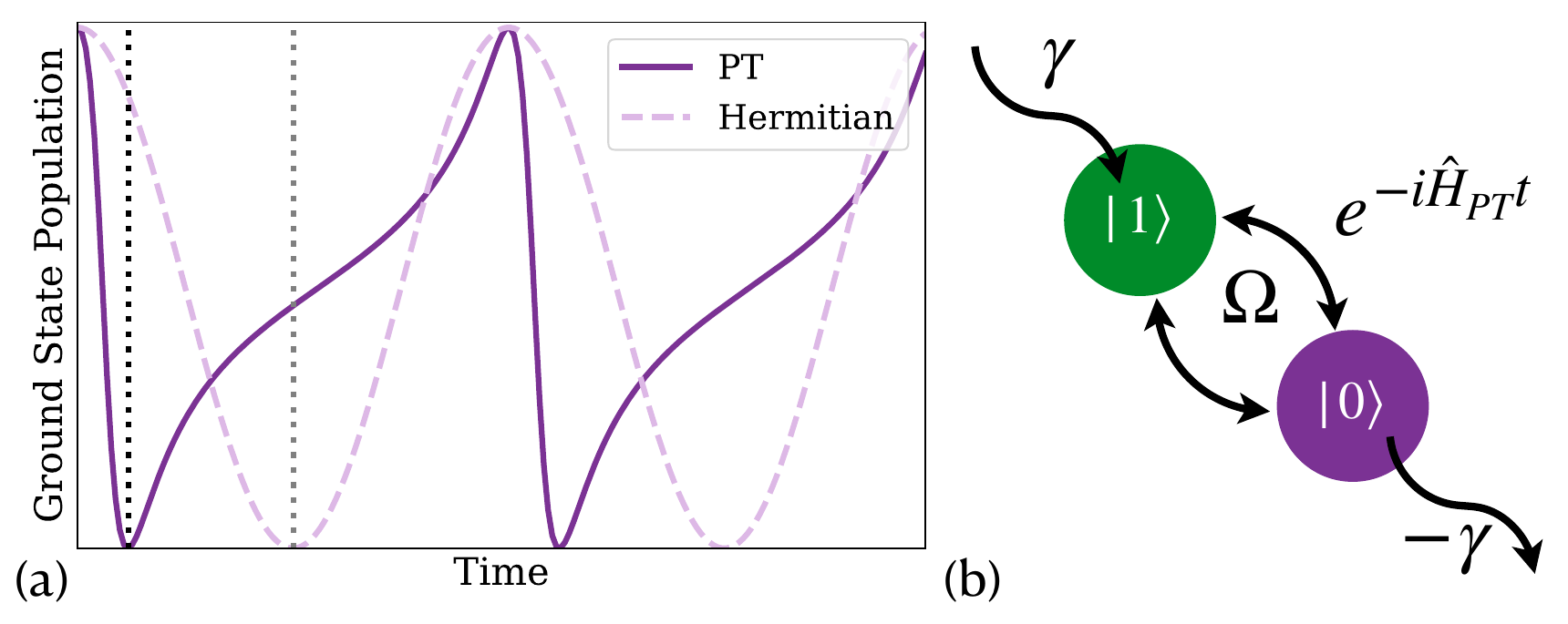}    
    \caption{
    (a) Comparison of the ground state population dynamics of a Hermitian (dashed line) and a $PT$-symmetric system (solid line). Dotted vertical lines show the time it takes for the ground state to evolve to the excited state for the $PT$ (black) and Hermitian dynamics (grey), demonstrating faster evolution on the $PT$ curve. (b) Schematic of a two-level system with balanced gain and loss where $\Omega$ is the coupling rate between the two levels and $\gamma$ is the gain and loss rate. }
\label{fig:ScehematicWdynamics}
\end{figure}

A particularly intriguing class of open quantum systems with non-Hermitian dynamics are Parity-Time ($PT$-) symmetric quantum systems~\cite{Bender1998}, which exhibit unique properties not observed in standard Hermitian quantum mechanics~\cite{Naghiloo2019, Abbasi2022, Chen2021, Chen2022, Wu2019, Holler2020, Guria2024, ZengLi2023, Liu2024, Quijandr2018, jzhang2023, kazmina2023, Wen2019, Dogra2021, Gnther2008, Kawabata2017, Bender2013, Lee2014, Chen2014, Wiersig2016, Liu2016, ElGanainy2018, Yoo2011, Rter2010, Klauck2019, Xiao2021, Li2019pt}. $PT$-symmetry refers to systems where the Hamiltonian remains invariant under combined parity- ($P$) and time-reversal ($T$) transformations. These systems possess real energy spectra in the so-called ``unbroken" $PT$-symmetric phase, despite the Hamiltonian being non-Hermitian. $PT$-symmetric systems possess a variety of interesting quantum properties, including the ability to tune the evolution time between two desired states such that the resulting time evolution is faster than a Hermitian system with the same spectrum~\cite{Bender2007, Gnther2008, Zheng2013}. This phenomenon is illustrated in Figure~\ref{fig:ScehematicWdynamics}(a), which compares the time evolution of the ground-state population under $PT$-symmetric (solid line) and Hermitian (dashed line) Hamiltonians with the same spectrum.

The real spectrum in the unbroken regime allows $PT$-symmetric systems to be mapped to an equivalent Hermitian form~\cite{Mostafazadeh2003}, permitting simulation with unitary quantum algorithms. Here we explore two approaches to simulating $PT$-symmetric systems on quantum computers by constructing an equivalent Hermitian Hamiltonian through similarity transformations. The similarity transformation allows for the offloading of the non-unitary dynamics to either classical pre/post-processing, or to additional quantum resources via dilation. Here, we explore both possible offloadings, referring to the algorithm which relies on classical postprocessing as the hybid algorithm, and the other the dilation algorithm. To examine the potential scaling of the hybrid algorithm, we extend beyond a single $PT$-symmetric system to consider two weakly interacting spins~\cite{Feyisa2024, Liu2024, LiZi2024, ZengLi2023}, where the weak interaction between the two $PT$-symmetric systems results in entanglement generation that is proposed to be faster than in Hermitian systems. We successfully simulate this entanglement generation by extending the one-qubit method via perturbation theory using a quantum simulator. Finally, we perform an error analysis on the single $PT$-symmetric algorithms to further assess their scalability. These results offer a scalable approach for simulating weakly coupled $PT$-symmetric qubits with unitary quantum algorithms.

\section{Theory and Methods}
\label{section:theory}
\subsection{Single $PT$-symmetric system}
We consider a two-level system with balanced gain and loss that can be effectively described by a $PT$-symmetric Hamiltonian,
\begin{equation}
    \hat{H}_{PT} = \begin{pmatrix}
        -i \gamma & \Omega\\
        \Omega & i \gamma    
    \end{pmatrix},
    \label{PT-Hamiltonian}
\end{equation}
where $\Omega$ is the coupling rate between the two states of the system and $\gamma$ is the rate of gain and loss, as depicted schematically in  Fig.~\ref{fig:ScehematicWdynamics}~(b). When $\Omega > \gamma$ this Hamiltonian is in the $PT$-symmetric unbroken phase and has real eigenvalues with non-orthogonal eigenstates. When $\Omega = \gamma$, the system reaches an exceptional point where eigenvalues are degenerate and the eigenstates coalesce. 

To explore the $PT$-symmetric region with real spectra, we investigate its Hermitian equivalent, which can be derived through a similarity transformation~\cite{Mostafazadeh2002, Mannheim2009, Zheng2022, Croke2015, Mostafazadeh2003, Wang2010}. In the unbroken regime $PT$-symmetric systems are a category of pseudo-Hermitian systems~\cite{Zheng2021,Mostafazadeh2002}, so there exists a Hermitian, invertible, linear operator $\eta$ such that $\hat{H}_{PT}^\dagger=\eta \hat{H}_{PT}\eta^{-1}$. This pseudo-Hermiticity allows for unitary evolution by redefining the inner product~\cite{Mostafazadeh2003, Mostafazadeh2002}. In the unbroken regime Eq.~\ref{PT-Hamiltonian} is parametrized as,
\begin{equation}
    \hat{H}_{PT} = \begin{pmatrix}
        -i \Omega \sin\alpha & \Omega \\
        \Omega & i \Omega \sin\alpha
    \end{pmatrix},
\end{equation}
where $\sin \alpha = \frac{\gamma}{\Omega}$. The corresponding eigenstates are,
\begin{align}
\label{eq:eigen1}
&|\psi_+\rangle = \frac{1}{\sqrt{2\cos\alpha}}(e^{\frac{i \alpha }{2}},e^{-\frac{i \alpha}{2}  })^T,\\
\notag
&|\psi_-\rangle = \frac{1}{\sqrt{2\cos\alpha}}(- e^{-\frac{i \alpha}{2}}, e^{\frac{i \alpha }{2}})^T.
\end{align}
The similarity transformation, $\eta$, which connects $\hat{H}_{PT}$ with $\hat{H}_{PT}^\dagger$ can be constructed from eigenstates of $\hat{H}_{PT}^\dagger$, which are complex conjugates of the original eigenstates, $|\phi_\pm\rangle = |\psi_\pm\rangle^*$. This similarity transformation takes the form,
\begin{align}
    &\eta =|\phi_+\rangle\langle\phi_+|+|\phi_-\rangle\langle\phi_-|\\
    &=\begin{pmatrix}
        \sec\alpha& i \tan\alpha\\
        -i\tan\alpha& \sec\alpha
    \end{pmatrix}.
\end{align}
Since $\eta$ is positive semidefinite in the unbroken regime we define its square root,~\cite{Mostafazadeh2003}
\begin{align}
    \sqrt{\eta} =\tau= \frac{1}{2}\left(
\begin{array}{cc}
 a+b & -i \left(a-b\right) \\
 i \left(a-b\right) & a+b \\
\end{array}
\right),
\end{align}
where $\tau$ is a positive-definite matrix, $a=\sqrt{\sec \alpha -\tan \alpha }$, and $b=\sqrt{\tan \alpha +\sec \alpha }$.
This matrix is used in a similarity transformation, resulting in a Hermitian equivalent of the $PT$-symmetric Hamiltonian, $\hat{h} = \tau \hat{H}_{PT} \tau^{-1}$, where $\hat{h}$ is Hermitian. In other words, $\hat{H}_{PT}$ is similar to $\hat{H}_{PT}^\dagger$ through $\eta$, and $\hat{H}_{PT}$ is similar to $\hat{h}$ through $\tau=\sqrt{\eta}$. Notably, for a time-independent Hamiltonian $\tau$ is a Hermitian matrix which does not depend on time. 
\begin{equation}
    \hat{h} = \left(
\begin{array}{cc}
 0 & \Omega \cos \alpha  \\
 \Omega \cos \alpha  & 0 \\
\end{array}
\right).
\end{equation}

The time-evolution operator for $\hat{H}_{PT}$ is,
\begin{equation}
    e^{-i \hat{H}_{PT}t}|\psi_0\rangle= e^{-it(\tau^{-1} \hat{h}\tau)}|\psi_0\rangle =\tau^{-1}e^{-i\hat{h}t}\tau|\psi_0\rangle.
    \label{eq:unitary_evo}
\end{equation}
This suggests that the simulation of the $PT$-symmetric dynamics can be done in three parts: initialization by $\tau\ket{\psi_0}$, unitary time evolution of the single qubit Hamiltonian $e^{-i\hat{h}t}$, and the final Hermitian operator $\tau^{-1}$ to complete the similarity transformation. We prepare the initial state with a one-qubit gate, $U_0$, that results in $\frac{1}{\mathcal{N}}\tau|\psi_0\rangle$ with normalization $\mathcal{N}$. After initialization, we apply the unitary time evolution with an $R_x(\theta)$ gate. For the final step, we implement $\tau^{-1}$ in one of two ways: first, using a one-qubit hybrid classical-quantum algorithm, and second by dilation and simulation in a two-qubit setting. We distinguish the two algorithms as either the hybrid or dilation algorithm. We show how the hybrid algorithm can be extended to weakly coupled $PT$-symmetric qubits via perturbation theory, without incurring significant additional classical costs.

\begin{figure}[h!]
    \centering
    \includegraphics[width = 1\columnwidth]{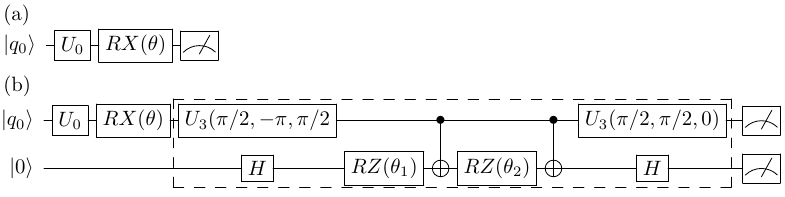}
    \caption{
        Quantum circuits corresponding to the (a) hybrid algorithm with $ e^{-i ht}\tau|0\rangle$ on a single qubit with tomography and (b) dilation algorithm with $\tau^{-1} \exp(-i ht)\tau|0\rangle$ using an ancillary qubit to implement the dilated $\tau^{-1}$.}
    \label{fig:circuits}
\end{figure}

In the hybrid approach, we perform the initialization and time propagation as shown in the circuit in Figure~\ref{fig:circuits}~(a), followed by tomography of the density matrix. We then multiply the $\tau^{-1}$ matrix classically to post-process the result and find the evolution of the $PT$-symmetric system in the initial basis. The classical multiplication of the matrix $\tau^{-1}$ requires the full tomography of the state, which significantly limits the scalability of this algorithm beyond NISQ devices. 

In the dilation approach, we realize the time evolution by implementing $\tau^{-1}$ as a dilated unitary, eliminating the need for full tomography of the density matrix to complete the similarity transform. Since $\tau^{-1}$ is non-unitary, we employ a unitary dilation of the operator, using the spectral decomposition of this Hermitian matrix, $\tau = UDU^\dagger$ and $\tau^{-1}= U^{*} D U^{T}$, where $U$ is unitary, and $D$ is diagonal.
The unitary dilation of $D$ results in a two-qubit unitary operator, $U_D$,
\begin{equation}\label{eq:dilation-block}
    U_{D} = \begin{pmatrix}
D_+ & 0 \\
0 & D_-
\end{pmatrix},
\end{equation}
where $D_{\pm}$ are diagonal matrices with entries,
\begin{equation}
    D_{\pm, jj} = D_{jj} \pm iD_{jj}\sqrt{\frac{1 - D_{jj}}{\Vert D_{jj} \Vert}},
    \label{eq:unit_norm_numbers}
\end{equation}
and $D_{jj}$ are diagonal elements of $D$. This diagonal operator can be implemented with combination of $R_z(\theta)$ and \textit{CNOT} gates as shown in Fig.~\ref{fig:circuits}~(b), where the angles are defined as,
\begin{align}
    &\theta_1=\frac{1}{2}\big(\textrm{arctan}(D_{\pm, 11}) + \textrm{arctan}(D_{\pm, 22})\big) \\
    &\theta_2=\frac{1}{2}\big(\textrm{arctan}(D_{\pm, 11}) - \textrm{arctan}(D_{\pm, 22})\big).    
\end{align}
The two-qubit algorithm results in the probablistic evolution of the state, conditioned on the ancilla being in the state $\ket{0}$ ; however, the dilation requires only one qubit, regardless of the size of the $PT$-symmetric Hamiltonian. Both optimal and efficient approximate algorithms are known for implementing unitary diagonal operators, although in this work we impement all diagonal operators exactly.~\cite{Welch2014}

\subsection{Two weakly-coupled $PT$-symmetric systems}
\label{theory-weakly-coupled}

Using first-order perturbation theory, we use the single-qubit eigenstates to simulate the dynamics of a two weakly-coupled $PT$-symmetric systems. We consider a weak interaction term of the form,
\begin{equation}
    \hat{H}_{\text{int}}=J(\sigma^1_+ \sigma^2_-+\sigma^1_-\sigma^2_+),
\end{equation}
where $\sigma_+ = (\sigma_x + i \sigma_y)/2$, $\sigma_- = (\sigma_x - i \sigma_y)/2$, and $\sigma_i$ are the Pauli operators.~\cite{Li2023,Liu2024} In this regime, the system retains a real spectrum within the perturbative approximation. The total Hamiltonian describing this system is given by,
\begin{equation}
    \hat{H}_{PT}^{12} = \hat{H}^1_{PT}\otimes \mathds{1}+ \mathds{1} \otimes \hat{H}^2_{PT}+\hat{H}_{\text{int}}^{12},
    \label{non-int2pt}
\end{equation}
where the superscripts list the qubit(s) on which the operator is supported. To find the eigenstates of a system of two non-interacting $PT$-symmetric systems, we can use the eigenstates of the one-qubit Hamiltonian from Eq.~\ref{eq:eigen1}. These new eigenstates are,
\begin{align}
    &|\psi_{--}\rangle = |\psi_{-}\rangle \otimes|\psi_{-}\rangle\\
    \notag
    &|\psi_{++}\rangle = |\psi_{+}\rangle \otimes |\psi_{+}\rangle\\
    \notag
    &|\psi_{+-}\rangle = |\psi_{+}\rangle \otimes |\psi_{-}\rangle\\
    \notag
    &|\psi_{-+}\rangle = |\psi_{-}\rangle \otimes |\psi_{+}\rangle.
\end{align}
Using first order perturbation theory we modify these states and generate a new similarity transformation to captures the dynamics of two weakly coupled $PT$-symmetric systems~\cite{Liu2024,ZengLi2023},
\begin{align}
    &\tau=\left\{\ket{\psi_{--}^{(1)}},\ket{\psi_{1}^{(1)}},\ket{\psi_{2}^{(1)}},\ket{\psi_{++}^{(1)}}\right\}
\end{align}
where the superscript (1) refers to first order perturbation and,
\begin{equation}
    \ket{\psi_{1,2}^{(1)}} = \ket{\psi_{-+}^{(1)}}\mp\ket{\psi_{+-}^{(1)}}.
\end{equation}
Using the identity $e^{-i\hat{H}^{12}_{PT}t}=e^{-i\tau E^{12} \tau^{-1}t}=\tau e^{-iE^{12}t}\tau^{-1}$, where $E^{12}$ is the diagonal matrix of the perturbed eigenvalues. We implement the evolution of this system using the hybrid algorithm approach introduced for the single $PT$-symmetric system. 

For a system of this size, we can measure the full density matrix via tomography, allowing calculation of the unique entanglement behavior of $PT$-symmetric systems near their exceptional points. We explore the entanglement dynamics using the concurrence, which is an entanglement measure for two qubit systems given by~\cite{Hill1997, Wooters1998},
\begin{equation}
    C(\rho) = \max\left(0, \lambda_1 - \lambda_2 - \lambda_3 - \lambda_4\right),
\end{equation}
where $\rho$ is the density matrix, and $\lambda_1, \lambda_2, \lambda_3$, and $\lambda_4$ are the square roots of the eigenvalues, in decreasing order, of the non-Hermitian matrix,
\begin{equation}
R = \rho \, \tilde{\rho}
\end{equation}
with
\begin{equation}
\tilde{\rho} = \left(\sigma_y \otimes \sigma_y\right) \rho^* \left(\sigma_y \otimes \sigma_y\right).
\end{equation}
Note that in the regime where the interaction is weak, the eigenvalues are real and the dynamics are now mapped on the dynamics of a Hermitian Hamiltonian evolution with unitary time operator. The similarity transformation, $\tau^{-1}$, is implemented on the initial state with single and two-qubit gates. Here, following the hybrid algorithm approach, the final $\tau$ is implemented as a classical post-processing; however future work could investigate the explicit decomposition of these circuits using the dilated three qubit algorithm. 

\section{Results}
\label{section:results}
\subsection{Single $PT$-symmetric system}

To demonstrate the two approaches for the time evolution of a single $PT$-symmetric system, we select parameters near the exceptional point, $\Omega = 7.5 ~\text{rad}/\mu s$ and $\gamma = 7  ~\mu s^{-1}$, which result in distorted oscillatory dynamics, similar to those in Fig.~\ref{fig:ScehematicWdynamics}~(a). We first identify the unitary operator $U_3(\theta,\phi,\lambda)$ that implements $\tau$ on the state $\ket{0}$. For the unitary time evolution, we note that $R_x(\theta) = e^{-i\hat{h}t}$, where $\theta =  \Omega t/2 \cos(\alpha)$. The derived quantum circuits are implemented on IBM-Sherbrooke quantum processing unit (QPU), as well as a simulator with the IBM-Sherbrooke noise model using $2^{12}$ shots~\cite{qiskit2024}. The device data were aquired over multiple days, with more details discussed in Appendix~\ref{sec:device-detail}
\begin{figure}[h!]
    \centering
    \includegraphics[width = 1\columnwidth]{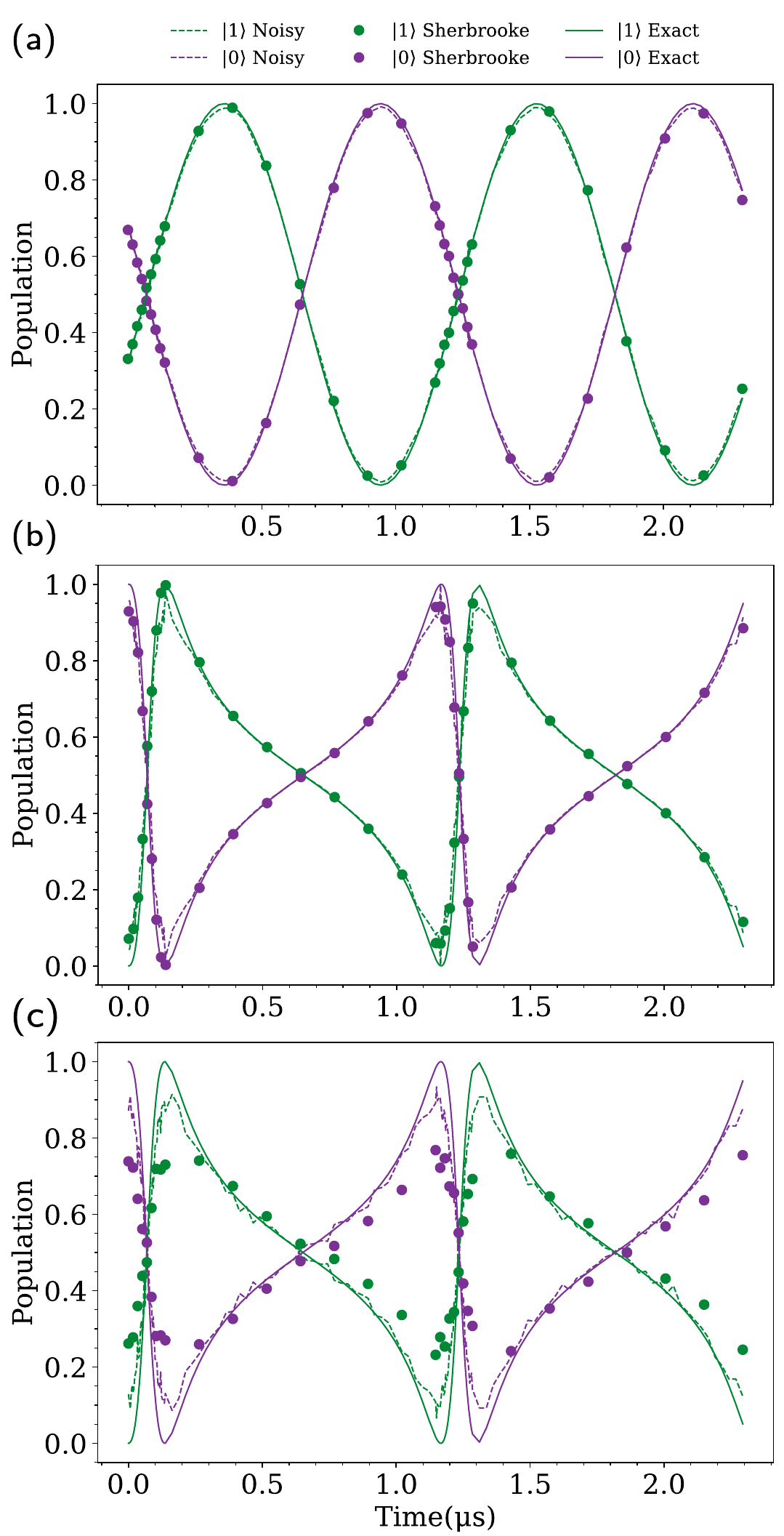}
    \caption{Ground (purple) and excited (green) state populations for the evolution of the system with (a) $e^{-i\hat{h}t}\tau|0\rangle$, (b) the one-qubit hybrid algorithm, and (c) the two-qubit dilation algorithm. Solid lines indicate the exact solutions, circles represent experimental data from the IBM Sherbrooke quantum processor, and dashed lines represent the noise-model data each with $2^{12}$ shots.}
    \label{fig:results1}
\end{figure}

The evolution of the ground- and excited-state populations on a single qubit initialized as $\tau\ket{0}$ are shown Figure~\ref{fig:results1}~(a) in purple and green respectively. The resulting evolution is a unitary evolution of a Hermitian Hamiltonian on a single qubit with the expected sinusoidal behavior. Figure~\ref{fig:results1}~(b) shows the ground- and excited-state populations from the hybrid algorithm considering the full density matrix measured by tomography in the Pauli basis then classically multiplied by $\tau^{-1}$ at each time step, resulting in a $PT$-symmetric time evolution. Both the QPU (circles) and the simulator results with the Sherbrooke noise model (dashed lines) agree well with the exact classical solution (solid lines). The dilation algorithm approach implements $\tau^{-1}$ with final post-selection on the $|0\rangle$ state of the ancillary qubit. The normalized populations of ground and excited states, $|00\rangle$ and $|01\rangle$ respectively, are shown in Fig.~\ref{fig:results1}~(c), with QPU results as circles. The QPU results are compared with IBM Sherbrooke noise model (dashed lines) as well as the exact classical solution (solid lines). This method remains NISQ-friendly as the ancillary qubit is used only for the sparse diagonal operator $\mathcal{\tilde{D}}$, as seen by the agreement of the QPU results with the classical solution.  This approach is naturally more sensitive to device noise due to the additional circuit volume; however, this approach can be applicable to coupled $PT$-symmetric qubits in the fault-tolerant regime via pertubation theory, because the diagonal operator can be implemented efficiently~\cite{Welch2014}.

\begin{figure}[h!]
    \centering
    \includegraphics[width = 1\columnwidth]{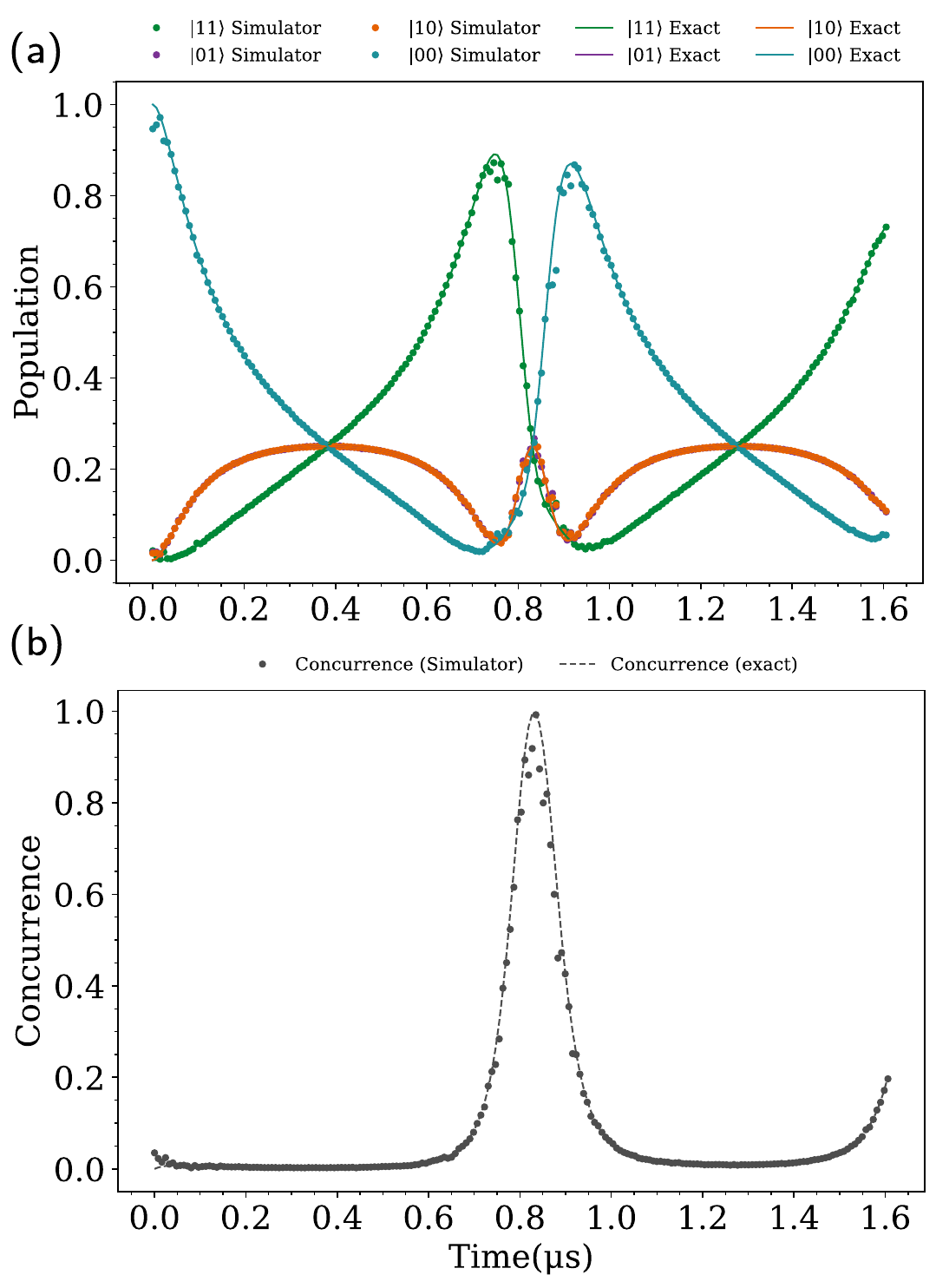}
    \caption{
       Time evolution of the (a) populations in the two-qubit basis states $|00\rangle$, $|01\rangle$, $|10\rangle$, and $|11\rangle$ over a duration of $1.6 \ \mu\text{s}$ and
(b) concurrence, which reaches its maximum when the populations of the basis states become approximately equal. }
    \label{fig:2qevolution}
\end{figure}

\subsection{Two weakly coupled $PT$-symmetric systems}

To assess the success of the perturbative treatment of two coupled $PT$-symmetric systems, we explore these dynamics with the parameters $\gamma=7 \mu s^{-1}$, $J=0.019 ~\text{rad}/\mu s$ and $\Omega=7.82 ~\text{rad}/\mu s$, and from an initially excited state, $\ket{11}$. We implement the first approach, the hybrid algorithm, from the single $PT$-symmetric system. Using the Aer simulator, we show the time evolution of the populations for two qubits in 
Figure~\ref{fig:2qevolution}~(a). Additionally, we compute the concurrence of this system using the density matrix of the evolved system, shown in Fig.~\ref{fig:2qevolution}~(b). This shows that the system achieves a maximally entangled state when all populations are equal. 

Noting here that we demonstrate the perturbative generalization using the first (hybrid) approach, but we could also use the second approach by diagonalizing $\tau_2^{-1}$ and dilating with an additional ancillary qubit. This method can be used to further simulate interaction Hamiltonians for multi-qubit systems with this symmetry~\cite{Feyisa2024, Liu2024}, and would allow us to work with larger systems since it no longer requires full state tomography.

\subsection{Single $PT$-symmetric Qubit Error Analysis}
\label{subsection:error}

\begin{figure}[h!]
    \centering
    \includegraphics[width=1\columnwidth]{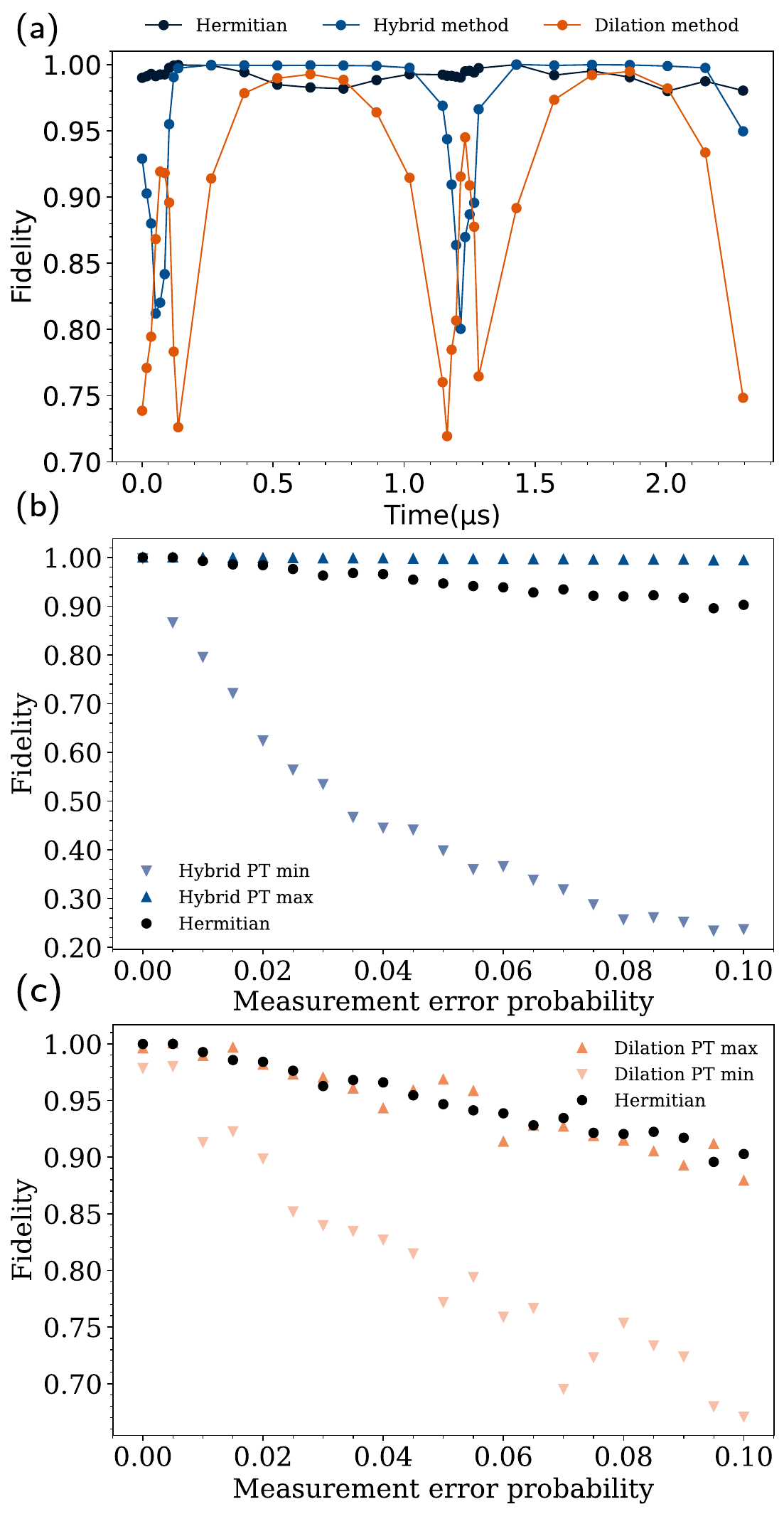}
    \caption{
        (a) The fidelity between the exact density matrix and the experimental results. Blue is the hybrid algorithm, orange shows the dilation algorithm results, and black shows the Hermitian evolution of a single qubit. (b) Simulated fidelity with respect to measurement error for the hybrid algorithm. (c) Simulated fidelity with respect to measurement error for the dilation algorithm.}
    \label{fig:results-error}
\end{figure}

Scaling these algorithms beyond two coupled $PT$ symmetric systems requires a thorough understanding of how the algorithms for a single $PT$-symmetric system respond to noisy conditions.  Returning to the single $PT$ symmetric system, Figure~\ref{fig:results-error}~(a) shows the fidelity between the experimental data and the exact solution for three cases: the Hermitian single-qubit evolution (black), the hybrid $PT$-symmetric algorithm (blue), and the dilation algorithm (orange). In the $PT$-symmetric single-qubit case, errors are amplified in the fast evolution regime, leading to a decrease in fidelity, whereas in the slow evolution regime, the fidelity remains stable. The difference in error behavior between Hermitian and $PT$-symmetric evolution arises due to classical post-processing, which can either suppress or amplify errors. In particular, matrix multiplication with the density matrix redistributes the error across its elements, so that, for example, $X$-type Pauli noise no longer corresponds to the same noise model after post-processing. In the dilation method, post-selection is applied independently at each time step, and the use of the $CNOT$ gate contributes to the error characteristics of this approach. 

To further assess the impact of noise on fidelity in each algorithm, we analyze the fidelity at two points in the dynamics under varying measurement error using a Qiskit noise model~\cite{qiskit2024}. From the results in Fig.~\ref{fig:results-error}~(a) for the hybrid algorithm, we choose time $t=1.22~\mu s$ where the fidelity is low and $t=1.77~\mu s$, where the fidelity is high. Here, the measurement error corresponds to the Pauli error $X$, where the state of the qubit flips with probability $P$~\cite{Nielsen:2010aa}. As shown in Fig.~\ref{fig:results-error}~(b) the fidelity at the minimum  from $t=1.22~\mu s$ (light blue down arrows) exhibits a non-linear dependence on measurement error, while it remains nearly constant at the maximum of the fidelity (dark blue up arrows) from $t=1.77~\mu s$. In contrast, the Hermitian Hamiltonian demonstrates a linear dependence on measurement error (black stars).

For the dilation approach depicted by the orange line in Fig.~\ref{fig:results-error}~(a), we observe a more complex behavior due to the presence of two-qubit measurement errors and entangling gates. In general the dilated algorithm is considerably less sensitive to the measurement error, with around 70\% fidelity at the highest rate, compared to 20\% for the hybrid algorithm. Examining the effect of measurement error on fidelity at different points in the absence of other errors, we find that at the maximum at $t=0.07 \ \mu \textit{s}$, the fidelity depends linearly on noise as shown in  Fig.~\ref{fig:results-error}~(c). However, at the minimum fidelity $t=0.01 \ \mu \textit{s}$, the dependence on measurement error increases. Because of increased quantum volume requirements, the dilated algorithm may incur larger error from two-qubit gate coherence limitations, which could explain the overall worse fidelity of the dilated algorithm; however, the classical post-processing in the hybrid algorithm is highly sensitive to the measurement error.

\section{Discussion and Conclusions}
\label{section:conclusions}
Here we have demonstrated that pseudo-Hermiticity can be harnessed to develop perturbatively generalizable quantum algorithms for $PT$-symmetric dynamics in the $PT$-unbroken phase. In the unbroken regime there exists a similarity transformation which allows the non-Hermitian Hamiltonian to be mapped onto a Hermitian Hamiltonian with the same spectrum. Using this similarity transformation, these dynamics can be simulated using both classical-quantum hybrid and ancilla-assisted algorithms. Both algorithms suggest the possibility to simulate $PT$-symmetric systems through real-time evolution by using the Hermitian-equivalent Hamiltonian rather than individual unitary gates. 

We then considered the evolution of two weakly interacting $PT$-symmetric systems to demostrate the algorithm extension to larger systems utilizing perturbation theory. Larger systems of equivalent weakly-interacting $PT$-symmetric qubits can further be simulated in this way. The weakly coupled qubits can be perturbatively treated in the basis of eigenstates of a single qubit, which allows for an efficient mapping of the coupled-qubit dynamics to a quantum algorithm. Here, we simulated the two-qubit dynamics using a hybrid algorithm, which effectively requires tomography of the density matrix. Alternatively, the coupled-qubit dynamics could be simulated with the fully quantum dilated algorithm, which would bypass the costly tomography.

Additionally, we analyzed the error of the hybrid and fully quantum algorithms to understand how they affect the fidelity of the $PT$-dynamics. By simulating varying measurement errors, we found that the hybrid algorithm is significnatly more sensitive to error, compared to the fully quantum algorithm. The largest errors occur during the fast part of the dynamics, and is absent from the Hermitian dynamics, highlighting the unique aspects of $PT$-symmetric systems which are potentially useful for quantum sensing applications.

Our results describe a mapping of weakly-coupled $PT$-symmetric qubits to a unitary gate-based quantum algorithm. The results suggest a scalable approach for simulting these dynamics on both near-term and fault-tolerant quantum computers. Error analysis highlights the unique challenges of simulating $PT$-symmetric systems on noisy quantum devices, which mirrors the unique behavior of $PT$-symmetric dynamics near exceptional points. This work opens an avenue for further investigation of non-Hermitian Hamiltonians on quantum computers. Future works will explore methods that enhances the efficiency of these algorithms as well as focusing on the potential application of these non-Hermitian systems. 

\noindent \textit{Acknowledgement.} We acknowledge K. W. Murch, W. Chen, J. R. Cruise, and A. Seidel for discussion. KHM acknowledges start-up funding from the University of Minnesota. This work was supported partially by the National Science Foundation through the University of Minnesota MRSEC under Award Number DMR-2011401.
\vspace{2pt}
\bibliographystyle{quantum}
\bibliography{main}
\appendix

\section{Device detail}
\label{sec:device-detail}
We used the default quantum state tomography in Qiskit Experiments version 0.8 for the hybrid algorithm, and we performed tomography with projective measurement with three tomographic gates on qubit 0 for the dilation algorithm. Both methods used samplerV2 with no error mitigation and suppression enabled. The QPU used here is ibm-sherbrooke, which is a 127 qubit quantum platform, where the processor type is specified as Eagle r3 superconducting qubit chip. The basis gates for this QPU are \{ECR, ID, RZ, SX, X\}. The median of respective T$_1$ and T$_2$ times are $\sim$ 267 $\mu$s, and $\sim$ 171 $\mu$s. The median error of each ID, SX and X gates  is $2.42 \times 10^{-4}$, whereas the error of RZ gate is 0. In terms of two-qubit gate error, the median of the ECR gate error is $8.02 \times 10^{-3}$. The median of the readout error is $1.66 \times 10^{-2}$, the state preparation and measurement (SPAM) error is $\sim$ 0.015.

\end{document}